\documentclass[%
reprint,
superscriptaddress,
aps,
longbibliography
]{revtex4-2}

\usepackage{graphicx} % Include figure files
\usepackage{amsfonts}
\usepackage{amssymb,amsmath}
\usepackage{bbm}
\usepackage{svg}
\usepackage{bm}% bold math
\usepackage{mhchem}
\usepackage[titletoc]{appendix}
\usepackage{dcolumn} % Align table columns on the decimal point
\usepackage{soul}                        % enable text highlight by \hl{}
\usepackage[normalem]{ulem}              % enable strikethrough by \sout{}
\newcommand{\xe}{\xi_\text{eff}}

\newcommand{\ket}[1]{|#1\rangle} % Ket
\renewcommand\vec{\mathbf}

\begin{document}

\title{On-Demand Zeeman Nuclear Frequency Comb Quantum Memory}

\author{Yanli Shi}
\affiliation{Institute for Quantum Science and Engineering, Department of Physics and Astronomy, Texas A\&M University, College Station, Texas 77843, USA}
\author{Xiwen Zhang}%
\email{xiwen@tamu.edu}
\affiliation{Institute for Quantum Science and Engineering, Department of Physics and Astronomy, Texas A\&M University, College Station, Texas 77843, USA}
\author{Yuri Shvyd’ko}
\affiliation{Advanced Photon Source, Argonne National Laboratory, Argonne, Illinois 60439, USA}
\affiliation{Institute for Quantum Science and Engineering, Department of Physics and Astronomy, Texas A\&M University, College Station, Texas 77843, USA}
\author{Olga Kocharovskaya}
\email{kochar@physics.tamu.edu}
\affiliation{Institute for Quantum Science and Engineering, Department of Physics and Astronomy, Texas A\&M University, College Station, Texas 77843, USA}

\date{\today}

\begin{abstract}
The emerging hard X-ray - nuclear interfaces offer unique potential advantages over traditional optical-atomic interfaces for room-temperature, solid-state quantum information processing, including lower background noise, tighter focusing, and exceptionally high resonance quality. Leveraging such interfaces, a major milestone was recently achieved with the first implementation of nuclear quantum memory in the hard X-ray range [S. Velten \textit{et al.}, Nuclear quantum memory for hard X-ray photon wave packets, Sci. Adv. \textbf{10}, eadn9825 (2024)] using the Doppler frequency comb protocol. However, this approach relies on the synchronous mechanical motion of multiple nuclear absorbers, posing experimental challenges for on-demand photon retrieval.
We propose an on-demand hard X-ray quantum memory based on reversing the direction of an external magnetic field in a single stationary solid-state nuclear absorber with sets of Zeeman sublevels. This scheme is exemplified by the quantum storage of an 1.41-$\mu$s single photon wave packet at 6.2 keV for over 10 $\mu$s in a $^{181}$Ta metallic foil, providing a feasible pathway for the first experimental demonstration of on-demand hard X-ray photon storage.
\end{abstract}

\maketitle

Quantum memories (QMs) are essential for quantum computing, long-distance quantum communication, and can also support on-demand single-photon sources, enhanced quantum sensing, and other applications~\cite{Bhaskar20Lukin, Lei23Hosseini}. They play a crucial role in synchronizing processes within an optical quantum network and enabling entanglement between nodes.
%~\cite{Lvovsky09Tittel, Zaiser16Wrachtrup, Simon17}
Although various QM techniques have been developed in the optical frequency regime using atomic ensembles~\cite{deRiedmatten08Gisin, Afzelius09Gisin, Hetet08Sellars, Zhang14Kocharovskaya, Serrano22Goldner}, the hard X-ray range remains largely unexplored despite several inherent advantages.
Due to their short wavelengths, X-rays can be focused to much tighter spots and penetrate deeper into many materials.
%Appel08Lvovsky, Moiseev11Tittel, Reim11Walmsley,
The orders of magnitude higher carrier frequencies result in much lower photon detection noise and offer a considerably broader available bandwidth.
M{\"o}ssbauer nuclei naturally serve as material hosts for hard X-ray wave packets due to the matching transition frequencies. Compared with atomic ensembles with electronic transitions, they provide a unique combination of much higher atomic number densities and longer coherence times even at room-temperature in solids.
Advances in coherence control of nuclei and X-ray quantum optics~\cite{Liao12Keitel,  Zhang13Svidzinsky, Vagizov14Kocharovskaya, Shwartz14Harris, Haber17Rohlsberger, Zhang19Kocharovskaya, Volkovich20Shwartz, Radeonychev20Kocharovskaya, Heeg21Evers, Khairulin22Kocharovskayaa, Trost23Chapman, Shvydko23Kolodziej, Kuznetsova24Kocharovskaya, Velten24Rohlsberger, Shakhmuratov24Vagizov, Khairulin25Radeonychev, Afanasev65Kagan, Gerdau85Hannon, Hannon68Trammell, Hannon69Trammell, Hannon99Trammell, Helisto91Katila, Kagan65Afanasev, Kagan66Afanasev, Kagan73Afanasev, Kagan79Kohn, Kagan99Kagan, Shvydko89Smirnov, ShvydKo91Hertrich, Shvydko92Smirnov, Shvydko93Smirnov, Shvydko94Gerdau, Shvydko96Schindelmann, Shvydko99Shvydko, Smirnov80Rudenko, Smirnov84Realo, vanBurck87Hannon} have opened up exciting opportunities for X-ray quantum information, underscoring the timeliness of exploring this domain.
%Glover12Hastings, Peik21Thirolf,

Among the various optical QM techniques, the atomic frequency comb has been widely used for the storage of optical photon wave packets in solids~\cite{deRiedmatten08Gisin, Afzelius09Gisin}.
%by creating a comb structure in the absorption spectrum.
This comb is typically formed through hole-burning on a broad inhomogeneous absorption profile, with atoms outside the comb teeth shelved to auxiliary levels using strong pulse trains. Achieving a high-quality comb with low residual absorption requires delicate experimental methods that greatly complicate the process.
The challenges become insurmountable in the hard X-ray regime due to the absence of sufficiently strong optical pumping fields.

Recently, a frequency comb memory protocol capable for hard X-ray photon quantum storage was proposed~\cite{Zhang19Kocharovskaya}.
Using the Doppler effect in multiple resonant, moving M{\"o}ssbauer absorbers with fixed velocity spacing, a high-quality frequency comb with no residual absorption is created in nuclear transitions without the need for optical pumping, auxiliary energy levels,
or inhomogeneous broadening~\cite{Zhang19Kocharovskaya}.
Very recently, this Doppler frequency comb has been experimentally demonstrated in $^{57}$Fe for storing $14.4$~keV hard X-ray photons at the single-photon level~\cite{Velten24Rohlsberger}.
However, the photon storage time by this Doppler nuclear frequency comb (DNFC), as with any frequency comb memory, is predetermined by the inverse comb spacing. It was previously suggested that on-demand QM could be achieved in DNFC by a rapid and simultaneous reversal of absorber velocities~\cite{Zhang19Kocharovskaya, Zhang16Zhang}, but achieving such fast and synchronized mechanical motion poses technical challenges for experimental realization.

In this work we propose an alternative approach for creating the nuclear frequency comb by using equidistant Zeeman splitting of the ground and the excited states of a nucleus in a stationary resonant absorber within a uniform magnetic field.
The resulting Zeeman nuclear frequency comb (ZNFC) significantly reduces system size and complexity by substituting numerous mechanical elements with a uniform magnetic field and reducing the number of absorbers to one, taking advantage of the internal degree of freedom of the nuclear transitions.
Furthermore, the last feature enables a simple realization of on-demand QM for hard X-ray photons by merely reversing the direction of the external magnetic field. This, in part, emulates a time-reversal operation, though it does not constitute a full time-reversal symmetry transformation.
Its feasibility was experimentally demonstrated by the observation of time-reversed quantum beat pattern in nuclear resonant Bragg scattering of synchrotron radiation from \textsuperscript{57}FeBO\textsubscript{3} as a result of rapid reversal of magnetic field~\cite{Shvydko95Ruter}. In the same publication it was also proposed to use this method for the observation of $\gamma$-ray photon echo.

\begin{figure}[hbt!]
    \includegraphics[width=\linewidth]{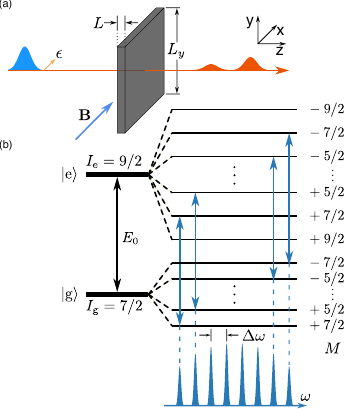}
    \caption{
    \label{fig:exp_setup_and_level_scheme}
    Illustration of the Zeeman nuclear frequency comb. (a) A thin M{\"o}ssbauer absorber of thickness $L \sim 1$~$\mu$m and transverse size $L_y\gg L$ in a uniform magnetic field $\vec{B}$ resonantly interacts with a linearly polarized hard X-ray photon, where the polarization vector $\vec{\epsilon}$ is along the quantization axis $x$.
    (b) The energy-level diagram of the $E_0 = 6.2$~keV nuclear E1 transition in $\alpha$-$^{181}$Ta metal under a uniform, static magnetic field $\vec{B}$. The $8$ transitions satisfying the selection rule $M_\text{g}-M_\text{e} = 0$ form a ZNFC in the absorption spectrum of the incident X-ray photon.
    Such a specific case of static ZNFC was realized experimentally by Chumakov \textit{et al.}~\cite{Chumakov95Chumakov} and published in~\cite{Leupold99Baron}. On-demand QM for the incident photon can be achieved by abruptly reversing the magnetic field direction at time $T_\text{sw}$, $\vec{B} \to -\vec{B}$.
    }
\end{figure}

A nuclear magnetic moment $\vec{\mu}$ in a uniform magnetic field $\vec{B}$ creates a Zeeman manifold with each sublevel shifted by a frequency $-\mu B (M/I) / h$, where $\mu = g \mu_\text{N} I$ is the magnetic dipole moment, $I$ is the nuclear spin quantum number, $M$ is the nuclear magnetic quantum number, $g$ is the nuclear $g$-factor, $\mu_\text{N}$ is the nuclear magneton, and $h$ is the Planck constant.
For simplicity, we assume that the nuclear ensemble resonantly interacts with an X-ray photon linearly polarized along the quantization axis as shown in Fig. \ref{fig:exp_setup_and_level_scheme} (a).
Such an in-plane magnetic field configuration also minimizes the induced eddy current decay time, if any (see later discussion).
Since the characteristic energy scale of the Zeeman splitting is set by $\mu_\text{N}=0.37$~mK/T, all ground-state energy sublevels remain nearly equally populated in moderate to low magnetic fields. Consequently, a ZNFC emerges in the absorption spectrum of the incident X-ray, with angular frequency given by
\begin{align}
\omega_{M_{<}} = \frac{1}{\hbar} M_< B \left(\frac{\mu_\text{g}}{I_\text{g}} - \frac{\mu_\text{e}}{I_\text{e}}\right), \label{Delta_omega}
\end{align}
where $M_<$ is the magnetic quantum number of the state (ground or excited) with the smaller spin, and subscripts ``$\text{g}$'' and ``$\text{e}$'' denote quantities associated with the ground and excited states, respectively.

Similarly to DNFC, photon echoes re-emit at times $nT_0$ after incident X-ray being absorbed by the ZNFC, where $n$ is an integer, $T_0 = 2\pi/\Delta \omega$ is the rephasing time, and $\Delta \omega = |\mu_\text{g}/I_\text{g} - \mu_\text{e}/I_\text{e}| B / \hbar$ is the angular frequency difference between adjacent comb teeth. The periodic re-emission arises from the beating of the discrete polarization waves established between sublevel pairs [such as illustrated in Fig. \ref{fig:exp_setup_and_level_scheme} (b)] with matching magnetic quantum numbers, allowing QM for hard X-ray photons with a predetermined storage time.
As follows from Eq. (\ref{Delta_omega}), simply reversing the magnetic field $\vec{B}$ to $-\vec{B}$ flips the ZNFC in the spectral domain due to the exchange of energies between magnetic sublevels $\ket{\pm M}$. 
The phase evolution of the nuclear polarization wave is thereby inverted, leading to construction interference and a controlled re-emission of the previously absorbed hard X-ray photon. 
In principle, the re-emission process following the magnetic field reversal could occur via all magnetic sublevel transitions allowed by the selection rules. However, only those channels that match the absorption pathways prior to the field reversal experience coherent enhancement~\cite{Shvydko93Smirnov}.
This approach enables on-demand quantum storage.

Since the number of teeth in a ZNFC is $2\text{min}(I_\text{g}, I_\text{e}) + 1$, nuclei with high spin quantum numbers in both ground and excited states are favored for achieving a larger memory bandwidth and higher storage fidelity.
Furthermore, $I_\text{g} \leqslant I_\text{e}$ is preferred to maximize utilization of the nuclear population.
However, nuclei with $I>1/2$ possess an electric quadrupole moment that couples to the electric field gradient (EFG) from the surrounding electrostatic environment, leading to energy shifts in the magnetic sublevels according to $M^2$ under axial symmetry and inducing level mixing in the general case~\cite{Cohen57Reif}.
As the interaction strength can range from less than a kilohertz to over a thousand megahertz, depending on the EFG, it typically strongly disrupts the ZNFC, compromising its ability to function as a quantum memory.
Fortunately, for nuclei positioned at sites with tetrahedral or higher symmetry, the EFG is necessarily zero, eliminating the electric quadrupolar interaction regardless of the quadrupole moments~\cite{Cohen57Reif}.

Therefore, nuclei with high spin numbers and large $g$-factor contrast between ground and excited states, residing at high-symmetry crystal sites in a uniform magnetic field, naturally form a ZNFC suitable for hard X-ray QM.

As a concrete example, we illustrate the implementation of this ZNFC-based QM in a Tantalum-181 metallic foil.
The choice of $^{181}$Ta, despite its significant photoioniaztion losses, stems from the unequivocal experimental evidence of ZNFC formation in previous studies~\cite{Leupold99Baron}, where a periodic echo sequence was observed in coherent forward scattering of synchrotron radiation.

Stable $\alpha$-tantalum metal has a body-centered cubic structure that eliminates quadrupolar interactions by symmetry and exhibits a high Lamb-M{\"o}ssbauer factor, $f_\text{LM}=0.96$~\cite{Leupold99Baron} even at room temperature. Its isotope $^{181}$Ta has a natural abundance of $99.988\%$. With nuclear spins $I_\text{g} = 7/2$ and $I_\text{e} = 9/2$, a uniform $\vec{B}$ induces an eight-tooth ZNFC in $^{181}$Ta metallic foil for a linearly polarized resonant photon at $6.2$~keV (see Fig.~\ref{fig:exp_setup_and_level_scheme}).
The comb teeth are spaced by $\Delta\omega /(2 \pi B) = (g_\text{e} - g_\text{g}) \mu_\text{N}/h = 3.781$~MHz~T$^{-1}$, where $g_\text{e} = 1.173$ and $g_\text{g} = 0.677$~\cite{Leupold99Baron, Wu05Wu}, each with a spectral width $\Gamma$.
For simplicity, we consider a lifetime-broadened transition, $\Gamma/(2\pi) = \Gamma_0/(2\pi) = 1/(2\pi T_1) = 18.23$~kHz, given the nuclear excited-state lifetime $T_1 = 8.73$~$\mu$s.
Small to moderate inhomogeneous broadening $\lesssim 4$ - $8 $ times the natural linewidth~\cite{Eibschutz83Disalvo, Chumakov95Wortmann, Leupold99Baron, Rohlsberger05Rohlsberger} increases due to crystal imperfection, limiting the comb teeth spacing and reducing the re-emitted photon amplitude. Nevertheless, echoes remain observable.
In fact, clear quantum beats from $^{181}$Ta ZNFC with an inhomogeneous broadening $\Gamma \sim 6\Gamma_0$ were reported~\cite{Chumakov95Chumakov,Leupold99Baron}, and the sharp peaks in the time dynamics were proposed as a means to probe external magnetic fields~\cite{Rohlsberger05Rohlsberger}.

\begin{figure}[hbt!]
    \includegraphics[width=\linewidth]{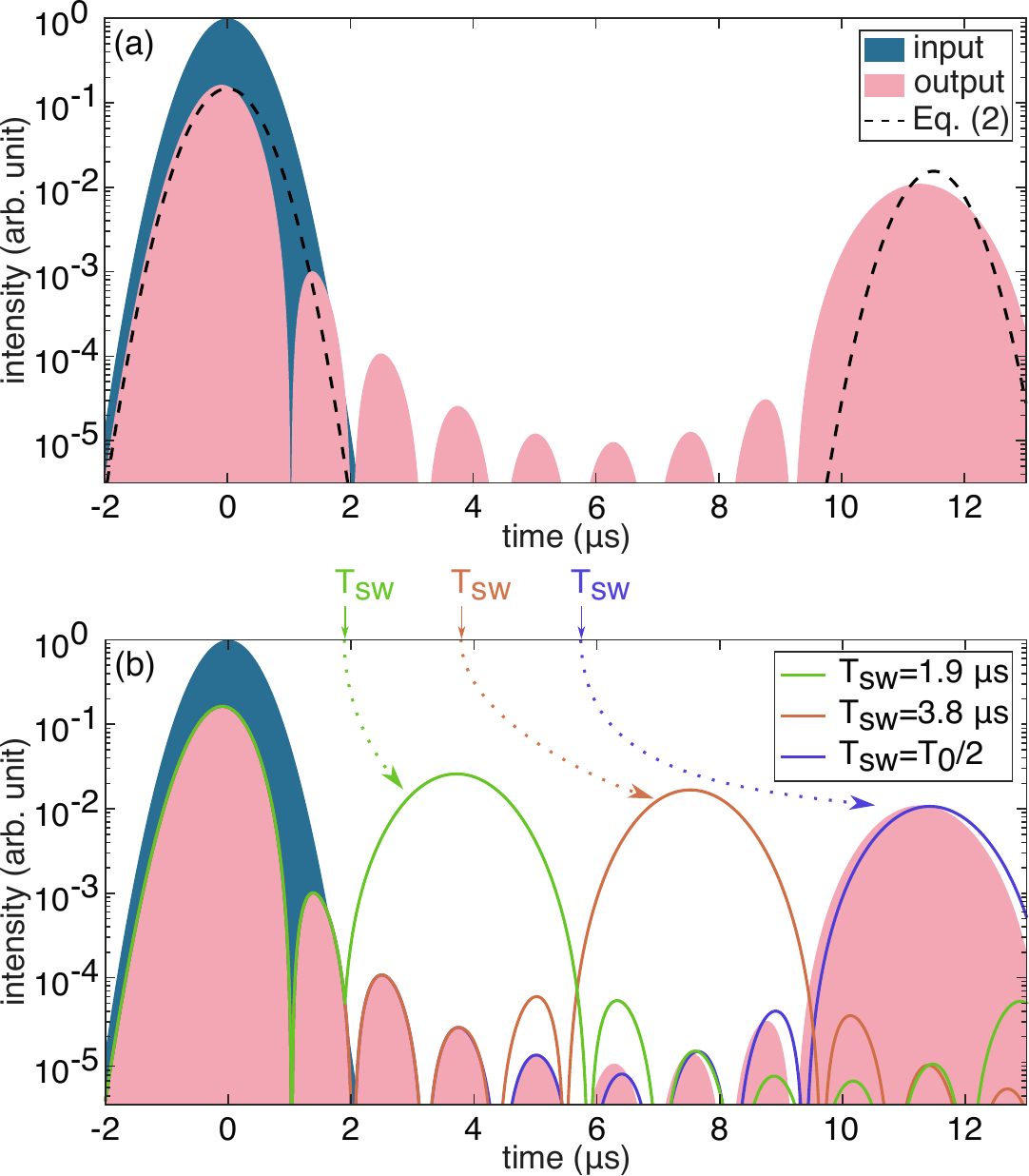}
    \caption{
    \label{fig:ZFC_ZQM_output}
    Zeeman nuclear frequency comb quantum memory in $^{181}$Ta.
    (a) Predetermined QM: an incident hard X-ray photon wave packet (navy blue, filled) is stored using $^{181}$Ta ZNFC without reversing the magnetic field direction. The re-emitted photon wave packet (light-red, filled) appears near the predetermined re-phasing time $T_0=11.50$~$\mu$s. Results from the numerical simulation (light-red, filled) of the Maxwell-Bloch equations agree well with the analytical prediction (black-dashed line) from Eqs. (\ref{eq:analytical_result}).
    (b) On-demand ZNFC QM: the retrieval of the stored photon is controlled by switching $\vec{B}$ to $-\vec{B}$ at an arbitrary time $T_\text{sw} \leqslant T_0$. The figure illustrates examples with switching times $T_\text{sw}=1.9$~$\mu$s, $3.8$~$\mu$s, and $T_0/2=5.75$~$\mu$s, as indicated by the color-coded arrows. The original predetermined re-emitted photon wave packet from (a), obtained without magnetic field switching, is included in the background for reference.
    In all cases, a field of $B=23$~mT generates the ZNFC in a $L=2.6$~$\mu$m-thick $^{181}$Ta metallic absorber with spectral spacing $\Delta \omega / (2 \pi) = 86.97$~kHz, finesse $\mathcal{F}=4.77$ and effective optical thickness per transition $\xi_\text{eff}^0=0.40$. The input resonant X-ray photon has a Gaussian temporal waveform with a FWHM field duration $\Delta t = 1.41$~$\mu$s.
    }
\end{figure}

Despite these prior realizations of ZNFC, its potential for on-demand hard X-ray QM has remain entirely unexplored.
Achieving viable quantum storage requires matching the spectral widths of the incident photon and the comb, along with implementing a magnetic field reversal to enable on-demand retrieval.
For instance, a resonant photon with a Gaussian temporal waveform and a full-width-at-half-maximum (FWHM) field duration of $1.41$~$\mu$s can be stored for a predetermined time of $T_0=2\pi/\Delta\omega=11.50$~$\mu$s using a ZNFC formed in a uniform magnetic field $23$~mT in $^{181}$ Ta, achieving a high fidelity of $91.20\%$, as numerically illustrated in Fig.~\ref{fig:ZFC_ZQM_output} (a).
On-demand retrieval can be achieved by rapidly switching the magnetic field to the opposite direction, as shown in ~\ref{fig:ZFC_ZQM_output} (b).

The light-matter interaction between the single-photon-level resonant hard X-ray field and a ZNFC is described by the Maxwell-Bloch equations~\cite{SM}. For a spectrally uniform nuclear frequency comb (NFC), X-ray photon echo in terms of Rabi frequency $\Omega$ without reversing the spectral components is
\begin{align}
\Omega(t,L)
&=
\beta e^{-\pi\xe^0/4}\Omega_\text{in}(\tau)
\nonumber\\
&\quad
- \beta \frac{\pi\xe^0}{2}e^{-\pi\xe^0/4}e^{-\pi/\mathcal{F}}\Omega_\text{in}(\tau-T_0)
\nonumber\\
&\quad
+ \text{higher sequence echoes},
\label{eq:analytical_result}
\end{align}
where $\Omega_\text{in}$ is the Rabi frequency of the incident X-ray pulse, $\mathcal{F}=\Delta\omega/\Gamma$ is the comb finesse, $\beta = \exp(-\mathcal{N}\sigma_\text{ph} L/2)$ is the off-resonant photon loss, $\xi_\text{eff}^0=\xi/(\mathcal{F}N)=\mathcal{N}\sigma_\text{R} f_\text{LM} L/(\mathcal{F}N)$ is the effective optical thickness per transition, $\xi$ is the total nuclear resonance optical thickness, $\mathcal{N}$ is the nuclear number density, $N=8$ is the total number of transitions, $\sigma_\text{ph}$ and $\sigma_\text{R}=1.1\times 10^{-18}$~cm$^2$ are off-resonant cross sections and resonant nuclear interaction cross sections, respectively, with internal conversion coefficient $70.5$~\cite{Leupold99Baron, Wu05Wu}, $\tau=t-L/c$ is the retarded time, $L$ is the absorber thickness along the X-ray propagation direction, and $c$ is the speed of light in vacuum.

\begin{figure}[hbt!]
    \includegraphics[width=\linewidth]{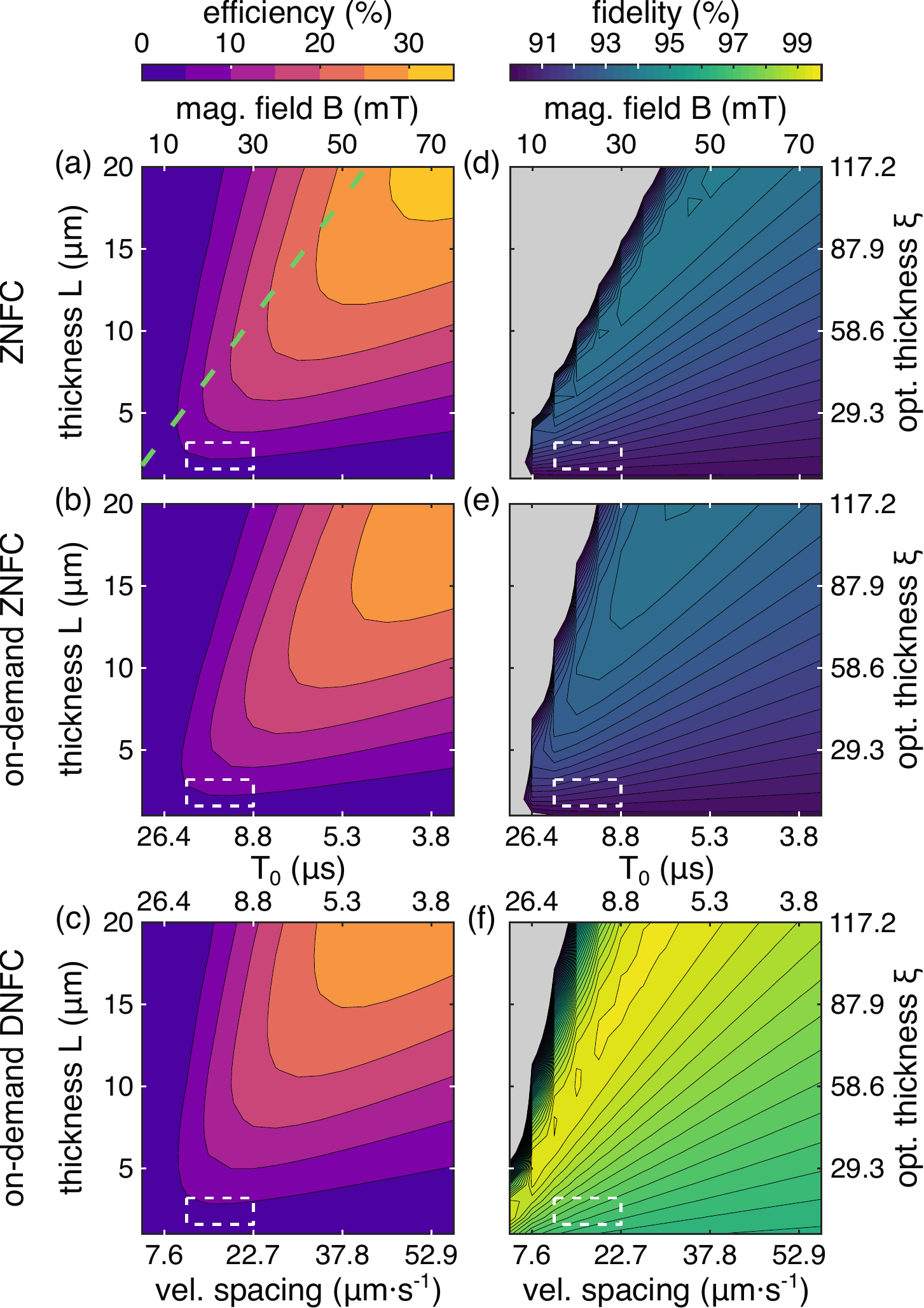}
    \caption{
    \label{fig:contours_PE0}
    Efficiencies (a-c) and fidelities (d-f) of hard X-ray quantum memories based on ZNFC and DNFC implemented in $^{181}$Ta, evaluated in the scenario without off-resonant photon loss to highlight the intrinsic performance of the scheme.
    Panels (a, d) show results for predetermined QM using ZNFC, while panels (b, e) and (c, f) present results for on-demand QM using ZNFC and DNFC, respectively.
    The DNFC scheme employs eight $^{181}$Ta absorbers, each with a thickness of $L/8$, moving at a velocity spacing $\Delta v = c \hbar\Delta\omega/E_0$.
    On-demand DNFC QM is realized by reversing the velocities of all absorbers at $T_\text{sw}$, leading to photon retrieval at $2T_\text{sw}$~\cite{Zhang19Kocharovskaya}.
    For a direct comparison, all switching times are set to $T_\text{sw} = T_0/2$, ensuring that photon echoes in all three cases emerge at $T_0$.
    The FWHM field duration of the incident single photon, with a Gaussian temporal waveform, is maintained at $\Delta t=8\ln(2)/(7\Delta\omega+\Gamma)$ to match the NFC bandwidths.
    The dashed green line in (a) corresponds to $\xi_\text{eff}^0 = 4/\pi$, and
    the white dashed boxes demark parameter ranges in Fig.~\ref{fig:contours_with_loss}.
    }
\end{figure}

In the absence of off-resonant photon loss, the maximum efficiency for storing a photon for a predetermined duration $T_0$ using a uniform NFC reaches $54\%$ at an effective optical thickness of $\xi_\text{eff}^0 = 4/\pi$~\cite{Zhang19Kocharovskaya, SM}, assuming a large number of comb teeth with negligible spectral width.
However, unlike a uniform NFC, the spectral weights in a ZNFC are proportional to the square of the Clebsch–Gordan coefficients of the appropriate transitions, resulting in a non-uniform optical thickness across the comb, with enhanced absorption in the central teeth. Consequently, when the incident hard X-ray field exhibits higher spectral flux near the central frequency --- such as in the case of a Gaussian spectral shape --- the comb is utilized more efficiently than a uniform one~\cite{Yeh19Liao}. By assigning less optical thickness to frequency regions with lower spectral weight, comparable efficiency can be achieved with reduced optimal total optical thickness.
Meanwhile, the comb finesse decreases with $\Delta \omega$, reducing storage efficiency for longer durations due to increased decoherence of the nuclear collective excitation.
This shifts the optimal $T_0$ to shorter times for a given optical thickness and further limits the achievable storage efficiency.
As a result, the optimal $\xi_\text{eff}^0$ decreases from $4/\pi$ to $1.1$, as shown in Fig.~\ref{fig:contours_PE0} (a) for the case of $\beta=1$.

\begin{figure}[hbt!]
    \includegraphics[width=\linewidth]{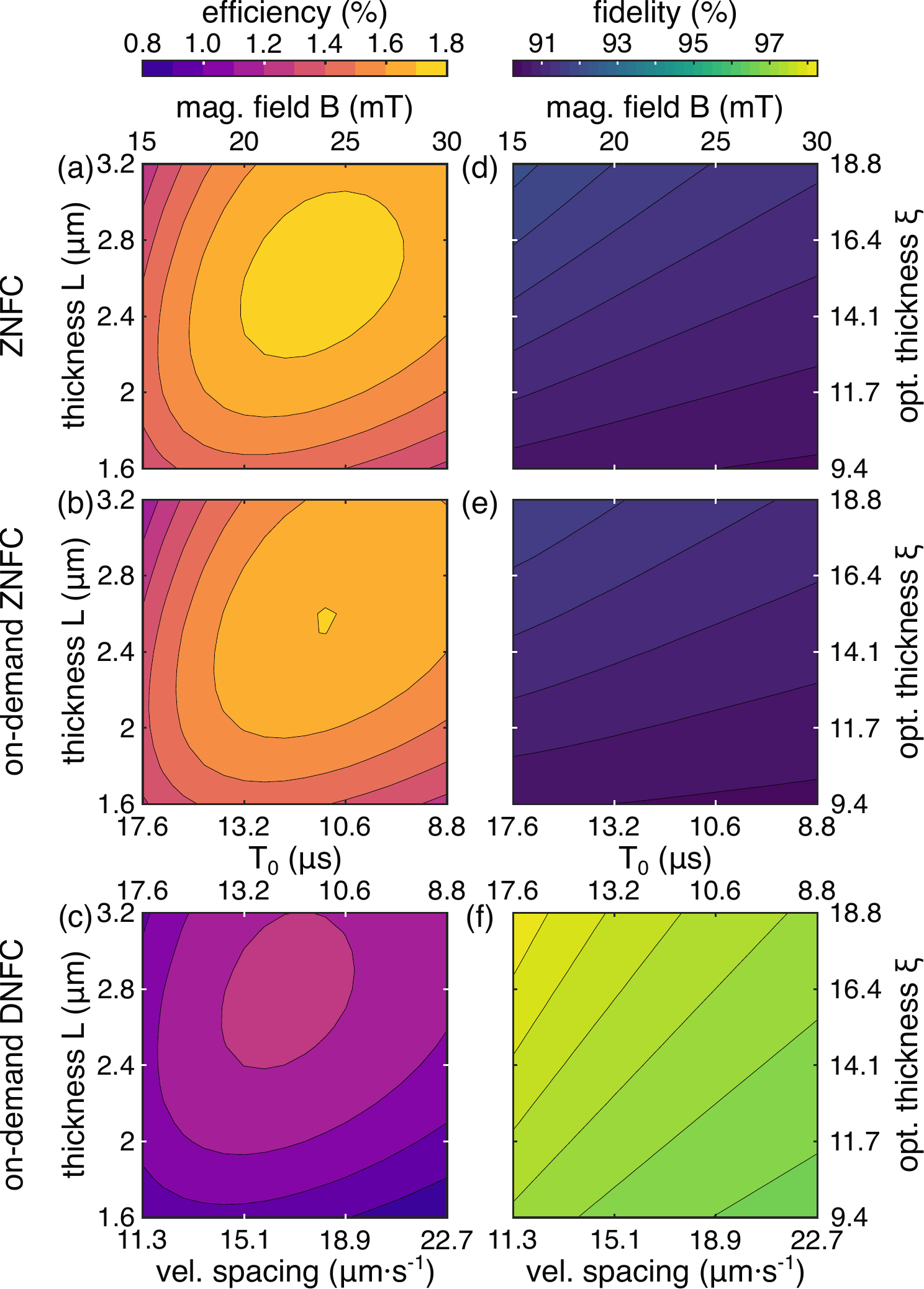}
    \caption{
    \label{fig:contours_with_loss}
    Efficiencies (a-c) and fidelities (d-f) of hard X-ray quantum memories based on ZNFC and DNFC implemented in $^{181}$Ta, accounting for realistic off-resonant photon loss, evaluated using the same parameters as in Fig.~\ref{fig:contours_PE0}.
    Unlike Fig.~\ref{fig:contours_PE0}, the storage efficiencies peak at $1.7\%$ for an optical thickness of $\xi \sim 15$ and a storage time of $T_0 \sim 10$~$\mu$s.
    }
\end{figure}

For a practical $^{181}$Ta ZNFC, off-resonant photon loss contributes significantly, constraining the optimal optical thickness and reducing the achievable experimental memory efficiency to $1.7\%$, as shown in Figs.~\ref{fig:ZFC_ZQM_output} (a) and~\ref{fig:contours_with_loss} (a).
This reduction arises because efficient resonant absorption of an incident X-ray pulse requires a relatively high nuclear optical thickness $\xi\gg1$. However, this also enhances off-resonant photo-absorption processes, since both resonant and off-resonant absorptions scale the same with their respective cross sections, $\sigma_\text{R}$ and $\sigma_\text{ph}$.
In $^{181}$Ta metallic foil, their ratio is $12.30$~\cite{Rohlsberger05Rohlsberger}.

On-demand QM is achieved by reversing the magnetic field direction at $T_{\rm sw}<T_0$, enabling the retrieval of the stored wave packet at $t=2T_{\rm sw}$ as illustrated in Fig.~\ref{fig:ZFC_ZQM_output} (b). Memory efficiency decreases with storage time due to decoherence.
 Compared with the ZNFC QM without magnetic field switching, the on-demand ZNFC allows flexible retrieval times up to $2T_0$ while maintaining similar storage efficiency and fidelity for a given storage duration, as shown in Fig.~\ref{fig:contours_with_loss} (a, d) and (b, e).
Figs.~\ref{fig:contours_PE0} (b) and ~\ref{fig:contours_with_loss} (b) show that the off-resonant photon loss affects the optimal optical thickness due to the previously explained mechanism, and imposes a strong limit on the maximum achievable memory efficiency, ideally close to $\sim 50\%$ as indicated by the numerical simulation in Fig.~\ref{fig:contours_PE0} (b).
Compared with on-demand DNFC QM, the on-demand ZNFC, implemented via magnetic field switching, attains similar memory efficiency [see Fig.~\ref{fig:contours_PE0} (c, f) and \ref{fig:contours_with_loss} (c, f)] while requiring only a single nuclear absorber and eliminating the need for mechanical modulation.

Experimentally, the switching speed of the magnetic field $B$ is constrained by both the external switching dynamics and the internal response time of the M{\"o}ssbauer absorber. The latter limitation arises from induced eddy currents, which oppose external switching.
For a slab geometry undergoing sudden modification of an applied magnetic field, the decay time of the induced eddy current is given by
 $(1/\pi^2) [L^2 L_y^2 / (L^2 + L_y^2)] (\mu_\text{r} \mu_0 / \rho) $,
where $\mu_\text{r}$ and $\mu_0$ are the relative and vacuum magnetic permeabilities, respectively, $\rho$ is the electrical resistivity, and $L_y$ is the transverse size of the absorber perpendicular to the external magnetic field~\cite{Bean59Nesbitt}.
In the specific configuration shown in Fig.~\ref{fig:exp_setup_and_level_scheme} (a),
the small thickness of $L\sim 2.5$~$\mu$m ensures that eddy currents in tantalum metal absorber decay rapidly within $\sim 6$~ps, imposing no practical limitation on the switching speed.
The external switching rate, in the case of an electromagnetic coil, is limited by the driving current, coil inductances, and eddy currents in surrounding conductive materials. In such a setup, a magnetic field of $\sim 10$~mT can be switched off within $\sim 1$ to a few microseconds, with potential improvements achievable through optimization of these limiting factors
~\cite{Kemmet11Weber, Eigen19Eigen, Kell21Gao}.

We have outlined a feasible approach for the pending experimental demonstration of on-demand hard X-ray QM using $^{181}$Ta ZNFC. In principle, the proposed protocol can be implemented in any high-symmetry M{\"o}ssbauer solid with high-spin states and suitable homogeneous and inhomogeneous broadenings.
A promising candidate is the $13.28$~keV nuclear transition in $^{73}$Ge crystal, with a lifetime of $4.2$~$\mu$s and nuclear spins of $I_\text{g}=9/2$ and $I_\text{e}=5/2$~\cite{Singh19Chen}.Combined with the corresponding ground and excited states $g$ factors, $g_\text{g}=-0.195$ and $g_\text{e}=-0.432$, this gives rise to a six-tooth ZNFC with a spectral spacing rate of $\Delta \omega / (2\pi B) =1.803$~MHz~T$^{-1}$. The ratio of resonant and off-resonant cross-sections is equal to 15. Thus, all the above parameters are similar to $^{181}$Ta. However, a nuclear transition in $^{73}$Ge has a higher polarity (E2). This cicumstance is very important, as it enables the suppression of off-resonant losses via the Borrmann effect \cite{Kuznetsova24Kocharovskaya, Batterman64Cole, Ruotsalainen16Huotari}, allowing for attainment of higher QM efficiency.
The other example is $^{45}$Sc~ isomer which emerged recently as the promising nuclear clock candidate~\cite{Shvydko23Kolodziej}.
Specifically attractive feature of $12.4$~keV nuclear transition of $^{45}$Sc is an exceptionally long lifetime of $0.47$ seconds, enabling ultra-long quantum storage.
However, implementing the DNFC protocol in the naturally broadened $^{45}$Sc isomer transition is enormously challenging, as it requires extremely slow motions of the order of $1$~nm~s$^{-1}$.
In contrast, the ZNFC approach can be formed in stationary cubic crystals such as scandium nitride~\cite{Steinadler22Brauniger, Al-Atabi20Edgar} and scandium trifluoride~\cite{Karimov19Popov} though with the tooth number limited by four due to the low value of the nuclear spin in the excited state $I_\text{e} = 3/2$. High polarity of the nuclear transition (M2) also allows using of the Borrmann effect for attaining the high efficiency. %The nuclear spin quantum numbers, $I_\text{g} = 7/2$ and $I_\text{e} = 3/2$, yield a four-tooth comb, which is nearly sufficient for storing wave packets with smooth temporal profiles, such as Gaussian-shaped pulses.
%Additional four teeth can be generated by incorporating a second stationary absorber under a slightly biased uniform magnetic field B' so that
%The ratio of resonant to off-resonant cross sections is only $\sim 5$ \cite{Shvydko23Kolodziej, Kuznetsova24Kocharovskaya, Rohlsberger2005Rohlsberger}, indicating more significant electron absorption losses than in $^{181}$Ta.%
%It is worth noting that the higher polarity of nuclear transition in $^{45}$Sc $^{45}$H (M2 instead of E1 in $^{181}$Ta) allows for suppression of off-resonant absorption losses via the Borrmann effect \cite{Kuznetsova24Kocharovskaya, Batterman1964Cole, Ruotsalainen2016Huotari} promising for achieving much higher efficiency.
Moreover, the higher polarity of the nuclear transition in $^{45}$Sc --- being M2 instead of E1 as in $^{181}$Ta --- enables the suppression of off-resonant absorption losses via the Borrmann effect~\cite{Kagan74Kagan, Kuznetsova24Kocharovskaya, Batterman64Cole, Ruotsalainen16Huotari}, allowing for attainment of higher QM efficiency.

Beyond the discussed hard X-ray - nuclear interface, the same protocol could potentially be implemented in atomic ensembles, such as rare-earth-doped crystals, for on-demand optical photon QM, provided that the inhomogeneous broadening of the optical electronic transitions is reduced well below the Zeeman splitting between adjacent sublevels.

The authors appreciate the financial support from the National Science Foundation (NSF, Grant No. PHY-240-97-34 ). The authors also gratefully acknowledge insightful discussions with J{\"o}rg Evers and Ralf R{\"o}hlsberger.

Data supporting the findings of this study are available from the corresponding author on request.

%O.K., X.Z. and Y.S. conceptualized the work; X.Z. and O.K. supervised the project; Y.S. and X.Z. developed the software; Y.S. performed the data curation and visualization; X.Z. evaluated the data; X.Z. and O.K. wrote the original manuscript; Y.S. and X.Z. wrote the original supplemental material; all authors reviewed and edited further versions of the manuscript.

%\bibliographystyle{apsrev4-2}
\bibliography{BibTex_QMTa181}

\end{document}